\documentclass[conference]{IEEEtran}
\IEEEoverridecommandlockouts
\usepackage{cite,url}
\usepackage{amsmath,amssymb,amsfonts}
\usepackage{graphicx}
\usepackage{textcomp}
\usepackage{xcolor}
\usepackage{color}
\usepackage{algorithm}
\usepackage[noend]{algpseudocode}
\usepackage{booktabs,makecell} 
\newcommand{\smallpara}[1]{\vspace{.6ex}\noindent{\bf #1}}
\newtheorem{example}{Example}%
\newtheorem{definition}{Definition}%
\usepackage{cancel}
\newcommand\nmodels{\mathbin{\cancel{\models}}}
\usepackage[symbol]{footmisc}

\newfloat{function}{htb}{}
\floatname{function}{Function}
\newfloat{procedure}{htb}{}
\floatname{procedure}{Procedure}
\usepackage[symbol]{footmisc}

\newtheorem{theorem}{Theorem}[section]

\newtheorem{lemma}[theorem]{Lemma}
\def\BibTeX{{\rm B\kern-.05em{\sc i\kern-.025em b}\kern-.08em
    T\kern-.1667em\lower.7ex\hbox{E}\kern-.125emX}}

\begin{document}

\title{{\sc FastAGEDs}: Fast Approximate Graph Entity Dependency Discovery}

\author{\IEEEauthorblockN{Guangtong Zhou\IEEEauthorrefmark{1}, 
Selasi Kwashie\IEEEauthorrefmark{2}, 
Yidi Zhang\IEEEauthorrefmark{1}, 
Michael Bewong\IEEEauthorrefmark{3}, 
Vincent M. Nofong\IEEEauthorrefmark{4},
Debo Cheng\IEEEauthorrefmark{5}, \\
Keqing He\IEEEauthorrefmark{6} and 
Zaiwen Feng\thanks{*Correspondence: Zaiwen.Feng@mail.hzau.edu.cn}\IEEEauthorrefmark{1}}
\IEEEauthorblockA{\IEEEauthorrefmark{1}\textit{College of Informatics,
Huazhong Agricultural University, Wuhan, Hubei, China}\\
\textit{Email: $\{$2019308210612, zhangyidi$\}$@webmail.hzau.edu.cn; Zaiwen.Feng@mail.hzau.edu.cn}
}
\IEEEauthorblockA{\IEEEauthorrefmark{2}\textit{AI \& Cyber Futures Institute, Charles Sturt University, 
Bathurst, Australia}\\
\textit{Email: skwashie@csu.edu.au}}
\IEEEauthorblockA{\IEEEauthorrefmark{3}\textit{School of Computing, Mathematics \& Engineering, Charles Sturt University, Wagga Wagga, Australia}\\
\textit{Email: mbewong@csu.edu.au}}
\IEEEauthorblockA{\IEEEauthorrefmark{4}\textit{Computer Science \& Engineering Department,
University of Mines \& Technology, Tarkwa, Ghana}\\
\textit{Email: vnofong@umat.edu.gh}}
\IEEEauthorblockA{\IEEEauthorrefmark{5}\textit{STEM,
University of South Australia, Adelaide, Australia}\\
\textit{Email: chedy055@mymail.unisa.edu.au}}
\IEEEauthorblockA{\IEEEauthorrefmark{6}\textit{School of Computing Science,
Wuhan University, Wuhan, Hubei, China}\\
\textit{Email: hekeqing@whu.edu.cn}}
}


\maketitle

\begin{abstract}
This paper studies the discovery of {\em approximate} rules in property
graphs. We propose a semantically meaningful measure of error for mining
graph entity dependencies (GEDs) at {\em almost} hold, to tolerate errors
and inconsistencies that exist in real-world graphs. 
We present a new characterisation of GED satisfaction, and devise a 
depth-first search strategy to traverse the search space of candidate 
rules efficiently. 
Further, we perform experiments to demonstrate the feasibility and 
scalability of our solution, {\sc FastAGEDs}, with three real-world
graphs.
\end{abstract}


\section{Introduction}
In recent years, researchers have proposed integrity constraints (e.g., 
{\em keys}~\cite{gkeys} and {\em functional dependencies} (FDs)~\cite{gfd}) 
for property graphs to specify various data semantics, and to address graph 
data quality and management issues. 
Graph entity dependencies (GEDs)~\cite{ged,gedj} represent a new set of 
fundamental constraints that unify keys and FDs for property graphs. 
A GED $\varphi$ over a property graph $G$ is a pair, $\varphi = (Q[\bar{u}], X\to Y)$, 
specifying the dependency $X\to Y$ over {\em homomorphic matches} of the graph pattern 
$Q[\bar{u}]$ in $G$. 
Intuitively, since graphs are schemaless, the graph pattern $Q[\bar{u}]$ identifies
the set of entities in $G$ on which the dependency $X\to Y$ should hold. 

GEDs have various practical applications in data quality and management 
(cf.~\cite{ged,gedj} for more details). For example, they have been extended 
for use in: fact checking in social media networks~\cite{fact_check}, entity 
resolution in graphs and relations~\cite{gdd,ontology}, consistency checking~\cite{incons}, 
amongst others. 

Like other data integrity constraints, the automatic discovery of GEDs is 
paramount for their adoption and practical use. Indeed, there is a growing
literature~\cite{gkeysminer,gfdminer,ged_disc} on the discovery problems of
GKeys~\cite{gkeysminer}, GFDs~\cite{gfdminer}, and GEDs~\cite{ged_disc}. 
However, these existing works focus on the discovery of dependencies that 
fully hold. Unfortunately, the existence of errors, exceptions and 
ambiguity in real-world data inhibits the discovery of semantically meaningful
and useful graph data rules.

Thus, in this paper, we investigate the automatic mining of GEDs that {\em almost} 
hold -- a well-suited rule discovery problem for real-world property graphs.
The main contributions are summarised as follows. First, we introduce and
formalise the approximate GED discovery problem via a new measure of error for 
GEDs based on its semantics. 
Second, we propose a novel and efficient algorithm, {\sc FastAGEDs}, to find 
approximate GEDs in large graphs. We introduce a new characterisation of GED
satisfaction by extending the notions of {\em disagree} and {\em necessary sets} 
to graph dependencies; and we develop an efficient depth-first search strategy
for effective traversal of the space of candidate GEDs, enabling fast discovery of
both full and approximate GEDs.
Third, we perform extensive experiments on real-world graphs to demonstrate
the feasibility and scalability of our discovery solution in large graphs.

\subsection{Related Works}
The discovery of approximate data dependencies is not new in the database 
and data mining communities, particularly, in the relational data setting.
Volumes of work exist on the discovery of approximate functional dependencies 
(FDs)~\cite{ge,diff,diff1,tane,afd,aifd,pyro} and its numerous extensions (e.g., 
conditional FDs~\cite{review,acfd,ucfd}, distance-based FDs~\cite{mfd, relaxed,
scamdd, ardd, cdd, add}, etc). 
The general goal is to find a reduced set of valid dependencies that (almost) 
hold over the given input data. However, this is a challenging and often 
intractable problem for most dependencies, including GEDs. 
The difficulties, in the GED case, arise due to four factors: a) the presence of
graph patterns as topological constraints; b) the LHS/RHS sets of GEDs 
have three possible literals; and c) the implication and validation analyses of 
GEDs are shown to be NP-complete and co-NP-complete(see~\cite{ged} for details); 
and d) the data may contain errors and inconsistencies.

\section{Preliminaries}
This section presents basic definitions and notions used throughout the
paper, a recall of the syntax and semantics of GEDs, as well as a 
discussion on the approximate satisfaction of dependencies in the 
literature.

\subsection{Basic Definitions and Notions}
The definitions of {\em property graph, graph pattern} and {match}
of graph patterns follow those in~\cite{ged,gedj}. 
We use alphabets $\mathbf{A, L, C}$ to be denote the countably infinite universal 
sets of {\em attributes, labels} and {\em constants} respectively.

\smallpara{Graph.}
We consider a directed property graph $G=(V,E,L,F_{A})$, where: (1) $V$ is a finite set 
of nodes; (2) $E$ is a finite set of edges, given by $E \subseteq V\times \mathbf{L}\times V$, 
in which $(v,l,v')$ is an edge from node $v$ to node $v'$ with label $l \in \mathbf{L}$;  
(3) each node $v \in V$ has a special attribute $id$ denoting its identity, and a label $L(v)$
drawn from $\mathbf{L}$; (4) every node $v$, has an associated list
$F_A(v)=[(A_{1},c_{1}),...,(A_{n},c_{n})]$ of attribute-value pairs, where 
$c_{i} \in \mathbf{C}$ is a constant, $A_{i} \in \mathbf{A}$ 
is an attribute of $v$, written as $v.A_{i}=c_{i}$, and $A_{i} \neq A_{j}$ if $i \neq j$.

\smallpara{Graph Pattern.}
A graph pattern, denoted by $Q[\bar{u}]$, is a directed graph $Q[\bar{u}]=(V_{Q}, E_{Q}, L_{Q})$,
where: (a) $V_{Q}$ and $E_{Q}$ represent the set of pattern nodes and pattern edges respectively;
(b) $L_{Q}$ is a label function that assigns a label to each node $v\in V_Q$ and each edge 
$e\in E_Q$; and (c) $\bar{u}$ is all the nodes, called (pattern) variables in $V_{Q}$.
All labels are drawn from $\mathbf{L}$, including the wildcard ``*" as a special label.
Two labels $l,l'\in \mathbf{L}$ are said to {\em match}, denoted $l\asymp l'$ iff: 
(a) $l=l'$; or (b) either $l$ or $l'$ is ``*". 

A {\bf match} of a graph pattern $Q[\bar{u}]$ in a graph $G$ is a homomorphism $h$ 
from $Q$ to $G$ such that: (a) for each node $v\in V_Q$, $L_Q(v) \asymp L(h(v))$;
and (b) each edge $e = (v,l,v') \in E_Q$, there exists an edge $e'=(h(v), l', h(v'))$
in $G$, such that $l\asymp l'$.
We denote the set of all matches of $Q[\bar{u}]$ in $G$ by $H(\bar{u})$\footnote[2]{simply 
$H$ when the context is clear}.
An example of a graph, graph patterns and their matches are presented below in 
Example~\ref{ex:graph}. 

\begin{figure}[t]
\centering
\includegraphics[width=\linewidth]{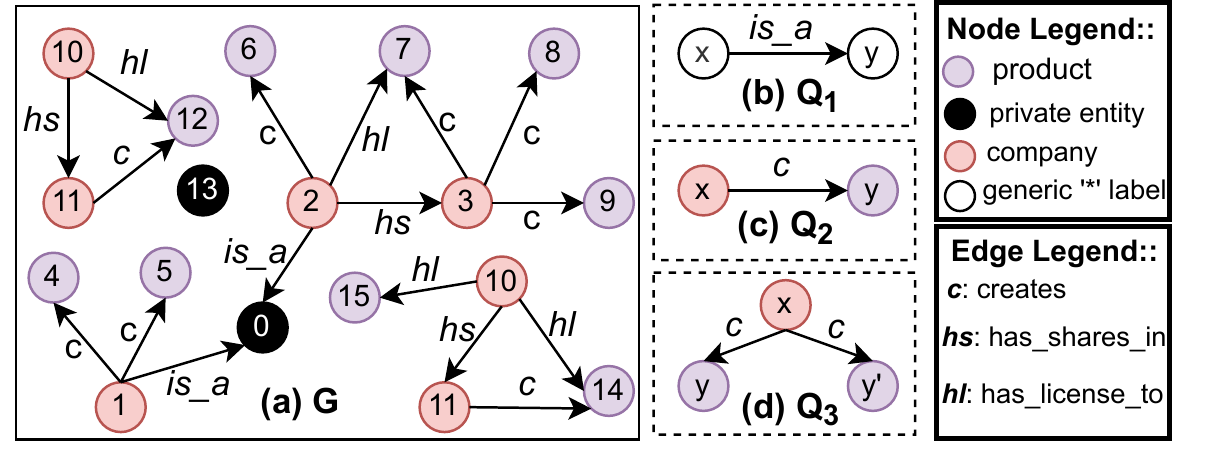}
\caption{Example: (a) graph; (b)--(d) graph patterns}
\label{fig:GED}
\end{figure}

\begin{example}[Graph, Graph Pattern \& Matches]\label{ex:graph}
Fig.~\ref{fig:GED}~(a) shows a simple graph; and in Fig.~\ref{fig:GED},~(b)--(d) 
are examples of three graph patterns. 
We present the semantics of each graph pattern and its corresponding homomorphic matches
in $G$ as follows: 
i) $Q_{1}[x,y]$ describes an ``\verb|is a|" relationship between two generic (``*" labelled)
nodes $x$ and $y$. The list of matches of this pattern in the example graph is 
$H_1(x,y)=[\{1, 0\}, \{2, 0\}]$.
ii) $Q_2[x,y]$ depicts a \verb|company| node $x$ with a \verb|create| relation with a \verb|product| 
node $y$, its matches in $G: H_2(x,y)=[\{1, 4\} \{1, 5\} \{2, 6\}, \{3, 7\}, \{3, 8\}, \{3,9\},
\{11, 12\}, \{11, 14\}]$.
iii) $Q_3[x,y,y']$ specifies a \verb|company| node $x$ with a \verb|create| relation with 
two \verb|product| nodes $y,y'$. Thus, matches in $G$ are $H_3(x,y,y') = [\{1,4,5\}, \{3,7,8\}, 
\{3,7,9\}, \{3,8,9\}]$.
$\square$
\end{example}

\subsection{Graph Entity Dependencies (GEDs)}
Here, we recall the definition of graph entity dependency (GED)~\cite{ged,gedj}.

\smallpara{GED Syntax. }
A GED $\varphi$ is a pair $(Q[\bar{u}],X \to Y)$, where $Q[\bar{u}]$ is a graph pattern, 
and $X,Y$ are two (possibly empty) sets of {\em literals} in $\bar{u}$. 
A literal $w$ of $\bar{u}$ is one of the following three constraints:
(a) $x.A = c$, (b) $x.A = y.A'$, (c) $x.id =y.id$, where
$x, y\in \bar{u}$, are pattern variables, $A,A' \in \mathbf{A}$, are non-$id$ attributes,
and $c\in \mathbf{C}$ is a constant. 
$Q[\bar{u}]$, and $X\to Y$ are referred to as the {\em pattern/scope} and {\em dependency} of 
$\varphi$ respectively. 

\smallpara{GED Semantics. }
Given a GED $\varphi=(Q[\bar{u}], X\to Y)$, a match $h(\bar{u})$ of $Q[\bar{u}]$ in $G$ 
{\em satisfies} a literal $w$ of $\bar{u}$, denoted by $h(\bar{u}) \models w$, if:
(a) when $w$ is $x.A = c$, then the attribute $h(x).A$ exists, and $h(x).A = c$; 
(b) when $w$ is $x.A = y.A'$, then the attributes $h(x).A$ and $h(y).A'$ exist
and have the same value; and (c) when $w$ is $x.id = y.id$, then $h(x).id = h(y).id$. 

A match $h(\bar{u})$ satisfies a set $X$ of literals if $h(\bar{u})$ satisfies every literal 
$w \in X$, (similarly defined for $h(\bar{u}) \models Y$).
We write $h(\bar{u}) \models X \to Y$ if $h(\bar{x}) \models X$ implies $h(\bar{x}) \models Y$.

A graph $G$ satisfies GED $\varphi$, denoted by $G \models \varphi$, if for all matches 
$h(\bar{u})$ of $Q[\bar{u}]$ in $G$, $h(\bar{x}) \models X \to Y$. 
A graph, $G$, satisfies a set $\Sigma$ of GEDs, denoted by $G \models \Sigma$, if for all 
$\varphi \in \Sigma$, $G \models \varphi$.

In Example~\ref{ex:ged}, we illustrate the semantics of GEDs with the sample graph and graph
patterns in Fig.~\ref{fig:GED}.

\begin{example}[Semantics of GEDs]\label{ex:ged}
We define exemplar GEDs over the sample graph in Fig.~\ref{ex:graph}(a), using the graph 
patterns in Figure~\ref{ex:graph}(b)--(d):\\
1) $\varphi_1: (Q_1[x,y], y.A=y.A\to x.A=y.A)$ -- this GED states that for any match $h(x,y)$ 
of the pattern $Q_1[x,y]$ in $G$ (i.e., $x$ \verb|is_a| $y$), if the node $h(y)$ has property
$A$, then $h(x)$ must have same values as $h(y)$ on $A$. \\
2) $\varphi_2: (Q_2[x,y], \emptyset \to y.creator=x.name)$ -- for every match $h(x,y)$ of $Q_2$ in $G$
(i.e., \verb|company| $h(x)$ \verb|create| \verb|product| $h(y)$), then $h(y.creator)$ and 
$h(x.name)$ must have the same value. \\
3) $\varphi_3: (Q_3[x,y,y'], \emptyset\to y.creator=y'.creator)$ -- this states for any match 
$h(x,y,y')$ in $G$ (i.e., the \verb|company| $h(x)$ \verb|create|
two products $h(y,y')$), then $y,y'$ must have same value on their property/attribute $creator$.
$\square$
\end{example}

\subsection{Approximate Satisfaction of Dependencies}
In the relational data model, there exists an extensive research on dependencies that 
almost hold due to their practical relevance in many real-world applications.
For instance, several measures have been proposed to quantify the extent of dependence
that exists between two sets of attributes/literals, using information theoretic 
ideas~\cite{ite1, ite2, ite3}, probabilistic~\cite{pe1} and non-probabilistic 
approaches~\cite{ge}. 
We refer interested readers to~\cite{comp_er} for a recent comparative study of
the subject.

This work attempts to present an intuitive, computationally efficient and 
effective measure of the degree of dependence for graph dependencies (GEDs, in 
particular).
Thus, we extend the definition of an error measure in~\cite{ge} based on the semantics
of graph dependencies for approximating dependencies in graphs.

\smallpara{Error Measure for GEDs.}
Consider the set $H(\bar{u})$ of matches of the graph pattern $Q[\bar{u}]$ in a graph $G$.
We say a match $h\in H$ {\em violates} a dependency $X\to Y$ over $Q[\Bar{u}]$ if: 
$h\models X$, but $h\nmodels Y$.
That is, intuitively, a dependency $X\to Y$ holds over $Q[\bar{u}]$ in $G$ if and only if 
there are no violating matches in $H$.
Hence we define an error measure $e_3$---{\em analogous to $g_3$ in~\cite{ge}}---to 
be the minimum number of matches to be eliminated to obtain
satisfaction.

Let $E_3$ be the number of matches that need to be eliminated from $H$ for $X\to Y$ to 
hold, expressed as:
\[
E_3(X\to Y, H) = |H| - max\{|J| \mid J\subseteq H, J\models X\to Y\}.
\]
The $e_3$ error measure for graph dependencies is given in Equations~\ref{eq:e3} as:

\begin{align}\label{eq:e3}
    e_3(X\to Y, H) = E_3/(|H|- |adom(X)|),
\end{align}
where $adom(X)$\footnote[3]{the max. $|adom(X)|$ of variable literals is \(2\), whereas that of 
constant literals is \(1\)} is the active domain of literals over $X$.
We remark that, the denominator of $e_3$ different from that of $g_3$ and based on the 
observation in~\cite{ite3} that the numerator is upper bounded -- consistent with
the revised $g'_3$ definition in~\cite{comp_er}.
$e_3$ ranges from $0$ to $1$; a value of \(0\) denotes full 
satisfaction of the dependency and a value of \(1\) shows an invalid dependency. 
Any value in between \(0\) and \(1\) indicates an approximate satisfaction of 
the dependency.

We define approximate GED {\em w.r.t.} $e_3$
and an error threshold $\epsilon\in [0,1]$ as follows in Definition~\ref{def:ap_ged}.

\begin{definition}[Approximate GED]\label{def:ap_ged}
    Let $\epsilon$ be a user-specified error threshold. 
    A GED $\sigma: (Q[\bar{u}], X\to Y)$ is an approximate GED in $G$, iff: 
    $e_3(X\to Y, H(\bar{u}))\leq \epsilon$.
    $\square$
\end{definition}

Suppose the pseudo-relational table in Fig.~\ref{fig:psuedo-table} represents
the attribute values of matches of pattern $Q_3$ in $G$ (from Fig.~\ref{fig:GED}).
We illustrate, in Example~\ref{ex:error}, examples of approximate GEDs 
based on $e_3$.

\begin{example}[Measuring Error of Graph Dependencies]\label{ex:error}
Consider the GEDs $\sigma_4: (Q_3[x,y,y'], X_1\to X_2)$, and $\sigma_5: (Q[x,y,y'], X_3 \to X_1)$,
where $X_1$ is $\{y.\verb|year|=y'.\verb|released|\}$, $X_2$ is $\{ y.\verb|name|=y'.\verb|name|\}$,
and $X_3$ is $\{x.\verb|name|=\verb|EA|\}$.
We compute $e_3(\sigma_4, H) = \frac{4 - 1}{4-2} = 0.5$ -- requires removal of only
match ($h_1$) and $|adom(X_1)|=2$; and $e_3(\sigma_5, H) = \frac{4-2}{4-1} = 0.667$
-- requires removal of two matches ($h_3, h_4$) and $|adom(X_3)|=1$.
    $\square$
\end{example}

\section{Problem Formulation}
The problem of mining GEDs is studied in~\cite{ged_disc}. In this section,
we recall relevant notions and definition on the discovery of GEDs, and 
extend the problem to the approximate satisfaction scenario.

Given a property graph $G$, the general GED discovery problem is to find a 
{\em minimal cover} set of GEDs that hold over $G$. Thus, intuitively, the
approximate GED discovery is to find a minimal cover set of GEDs that 
almost hold, based on a given error measure and error threshold.
In the following, we formulate and present a more formal definition of the 
problem. 

\subsection{Persistent, Reduced, and Minimal GEDs}
In rule discovery, it is paramount to return a succint and non-redundant
set of rules. Thus, in line with the data dependency discovery 
literature~\cite{ged_disc, gfd_disc, fd_rev, relax}, we are interested in
a cover set of {\em non-trivial} and {\em non-redundant} dependencies over 
{\em persistent} graph patterns.

\smallpara{Persistent Graph Patterns. }
Let $M=\{m_1, \cdots, m_k\}$ be the set of isomorphisms of a pattern 
$Q[\bar{u}]$ to a graph $G$; and $D(v) = \{m_1(v), \cdots, m_k(v)\}$ be 
the set containing the distinct nodes in $G$ whose functions $m_1,\cdots, 
m_k$ map a node $v\in V$. 

The minimum image based support~\footnote[4]{we adopt this metric due to
its computational efficiency and anti-monotonic property}
(MNI)~\cite{mni} of $Q$ in $G$,
denoted by ${\sc mni}(Q, G)$, is defined as:
\[{\sc mni}(Q, G) = min\{x\mid x =|D(v)|,\mbox{ } \forall \mbox{ }v \in  V \}.\]

We say a graph pattern $Q[\bar{u}]$ is {\em persistent} (i.e., {\em frequent}) 
in a graph $G$, if ${\sc mni}(Q, G) \geq \tau$, where $\tau\in \mathbb{N}$ is 
a user-specified minimum MNI threshold.

\smallpara{Trivial GEDs. } 
We say a GED $\varphi: (Q[\bar{u}], X\to Y)$ is 
{\em trivial} if: (a) the set of literals in $X$ cannot be satisfied (i.e., $X$ 
evaluates to {\tt false}); or (b) $Y$ is derivable from $X$ (i.e., $\forall 
\mbox{ }w\in Y$, $w$ can be derived from $X$ by transitivity of the equality operator). 
We are interested in mining only non-trivial GEDs.

\smallpara{Reduced GEDs. }
Given two patterns $Q[\bar{u}]=(V_Q,E_Q,L_Q)$ and $Q'[\bar{u'}]= (V'_{Q'},E'_{Q'}, 
L'_{Q'})$, $Q[\bar{u}]$ is said to {\em reduce} $Q'[\bar{u'}]$, denoted 
as $Q\ll Q'$ if: (a) $V_Q\subseteq V'_{Q'}$, $E_Q\subseteq E'_{Q'}$; or 
(b) $L_Q$ upgrades some labels in $L'_{Q'}$ to wildcards. 
That is, $Q$ is a less restrictive topological constraint than $Q'$.

Given two GEDs, $\varphi = (Q[\bar{u}], X\to w)$ and $\varphi' = (Q'[\bar{u'}],
X'\to w')$. $\varphi$ is said to {\em reduce} $\varphi'$, denoted by 
$\varphi\ll \varphi'$, if: (a) $Q\ll Q'$; and (b) $X\subseteq X'$ and $w=w'$.

Thus, we say a GED $\varphi$ is {\bf reduced} in a graph $G$ if: 
(a) $G\models \varphi$; and for any $\varphi'$ such that $\varphi'\ll 
\varphi$, $G\not\models\varphi'$.
We say a GED $\varphi$ is {\bf minimal} if it is both non-trivial and reduced. 

\begin{figure}[!t]
    \centering
    \includegraphics[width=0.48\textwidth]{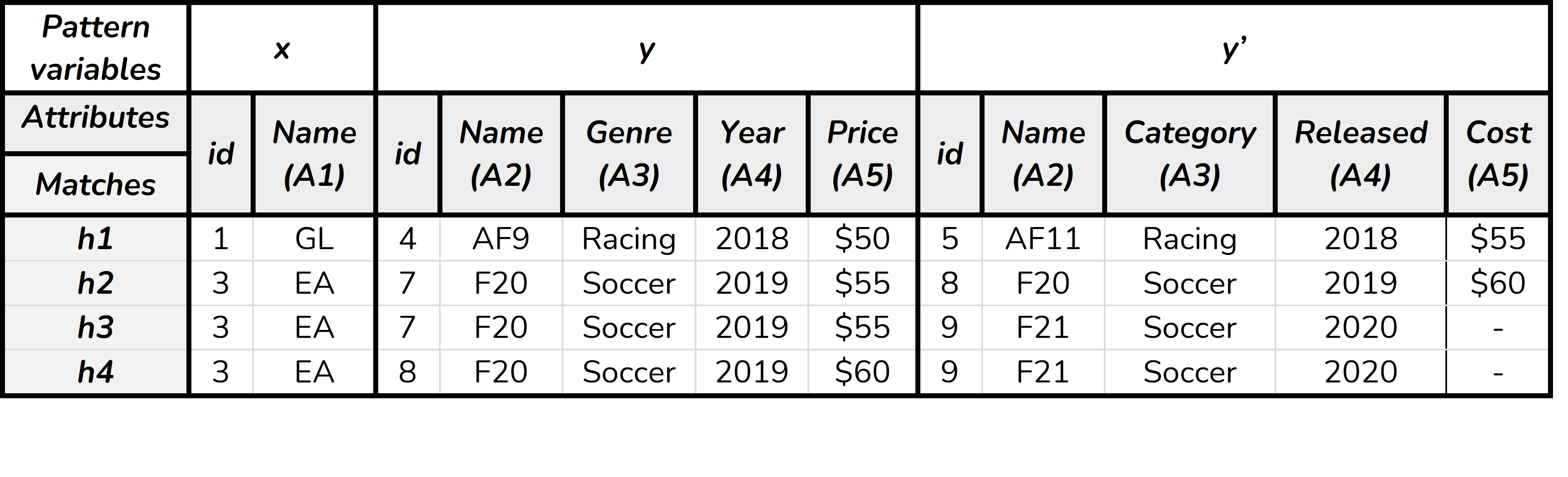}
    \vspace{-10pt}
    \caption{A pesudo-relation of matches of $Q_3$ in $G$ (cf.~Fig.~\ref{fig:GED})}
    \label{fig:psuedo-table}
\end{figure}

\subsection{Problem Definition}
In this paper, we propose and study the approximate GED discovery problem.

\begin{definition}[Approximate GED Discovery Problem]\label{def:aged_disc}
Given a property graph, $G$, a user-specified MNI threshold, $\tau \in \mathbb{N}$,
and $e_3$ error threshold, $\epsilon \in [0,1]$. The approximate GED discovery is
to find a set $\Sigma$ of all minimal GEDs in $G$, such that for each 
$\sigma: (Q[\bar{u}], X\to Y)\in \Sigma$, we have: a) ${\sc mni}(Q, G)\geq \tau$
and b) $e(X\to Y, H)\leq \epsilon$.
$\square$
\end{definition}

We present our approximate GED discovery solution, {\sc FastAGEDs}, in
the following section.

\section{Fast Approximate GED Mining}
A GED specifies attribute dependencies over topological patterns in a
graph. Thus, we first, examine the discovery of the graph patterns 
(subsection~\ref{ssec:gpat}) which shall serve as the ``scope" for the 
attribute dependencies discovery (subsection~\ref{ssec:dep}).
A simplified pseudo-code of our solution is presented in Algorithm~\ref{alg:fast}.

\subsection{Graph Pattern (Scope) Discovery}\label{ssec:gpat}
Here, we present a brief discussion on the discovery of frequent
graph patterns as the ``scopes" or loose-schema over which we mine
the attribute dependencies. 
We do not claim novelty of contribution on this task -- which 
is a well-studied topic in the data mining and database 
literature~\cite{grami, prox, empty, arab}. 

In this work, we adopt the efficient MNI-based technique, 
{\sc GraMi}~\cite{grami}, to find $\tau$-frequent graph patterns as the 
scopes for our dependencies (line~\ref{l:mfp} of Algorithm~\ref{alg:fast}); 
and employ the pruning strategies in~\cite{ged_disc} to return a set of reduced graph patterns (line~\ref{l:rrp}).
We then mine dependencies over the set of matches of each reduced graph 
patterns in the following.

\subsection{Dependency Mining}\label{ssec:dep}
Let $\mathcal{Q}$ be the set of reduced $\tau$-frequent graph patterns
in a property graph $G$. We mine a minimal cover set, $\Sigma(Q, G, \epsilon)$, 
of approximate GEDs for each pattern $Q\in \mathcal{Q}$ in $G$ based on a
user-specified minimum threshold $\epsilon$ in $G$ (lines~\ref{l:md1}--\ref{l:md2}
of Algorithm~\ref{alg:fast}).

\smallpara{A Novel GED Characterisation. }
We present a novel characterisation of GEDs satisfaction based on the concept of
{\em disagree set}, first introduced in~\cite{diff,diff1} for relational
functional dependency discovery. 

Let $H(Q[\bar{u}], G)$ be the homomorphic matches of $Q[\bar{u}]\in \mathcal{Q}$ 
in the input graph $G$. We introduce the concepts of {\em item/itemset}, to map
onto constant, variable, and \verb|id| literals over the attributes of 
pattern variables $x,y\in \bar{u}$ in $Q$ as follows.

\begin{definition}[Item, Itemset, \& Itemset Satisfaction]\label{def:item}
    Let an item-name, $\alpha$, be an attribute of a pattern variable $x\in \bar{u}$
    of a graph pattern $Q$. An {\em item} is a triple, consisting of item-name, pattern
    variable(s) $u\subseteq \bar{u}$, and constant $c \in \mathbf{C}$, 
    denoted by $\alpha[u;c]$.
    An {\em itemset}, $X=\{\alpha_1[u_1;c_1], \cdots, \alpha_n[u_n;c_n] \mid 
    \forall \mbox{ } i, j \in [1,n], i\neq j\}$, 
    is a set of items with unique item-names. 
    We say a match $h\in H$ satisfies an itemset $X$ iff: $h(u_i) \models 
    \alpha_i[u_i;c_i], \forall \mbox{ } \alpha_i[u_i;c_i]\in X$. $\square$
\end{definition}

An item $\alpha[u;c]$ over $\bar{u}$ of $Q$ uniquely maps to constant, variable,
and \verb|id| literals of $w$ of $\bar{u}$, respectively, as follows:
\begin{itemize}
    \item when $w$ is $x.A=c$, then its item is $\alpha[x;c]$
    \item when $w$ is $x.A_1 = y.A_2$, then its item is $\alpha_{12}[x,y;\verb|_|]$
    \item when $w$ is $x.\verb|id|=y.\verb|id|$, then its item is 
    $\verb|id|[x,y;\verb|_|]$.
\end{itemize}
Thus, given a graph pattern $Q[\bar{u}] \in \mathcal{Q}$, we define constant
items for all attributes of pattern variables $x \in \bar{u}$, and variable
item over pattern variable pairs of the same kind $x,y \in \bar{u}$.


\begin{definition}[Disagree and Necessary Sets]\label{def:diff}
Given the set of matches $H$ of $Q[\bar{u}]$ in $G$:
\begin{itemize}
    \item the {\em disagree set}, $D(h)$, of a match $h \in H$ is given by
    \[D(h) = \{\alpha[u;c] \mid h\nmodels \alpha[u;c]\}.\]
    \item the {\em necessary set} of an item, $\alpha[u;c]$, is defined as
    \[nec(\alpha[u;c]) = \{D(h)\setminus \alpha[u;c] \mid h\in H, \alpha[u;c]\in D(h)\}.\] 
\end{itemize} 
\end{definition}

Let $\mathcal{I}$ denote the set of all items over a graph pattern $Q[\bar{u}]$, and
$P(\mathcal{I})$ be the power set of $\mathcal{I}$. 
Given $M\subseteq \mathcal{I}, X\subseteq P(\mathcal{I})$, the set $M$ {\bf covers} $X$
if and only if, for all $Y \in X$, we have $Y\cap M\neq \emptyset$.
Furthermore, $M$ is a minimal cover for $X$ if there exists no $M'\subset M$ which
covers $X$. 

\begin{algorithm}[!t]
\caption{{\sc FastAGEDs}($G, \tau, \epsilon$)}
\label{alg:fast}
\begin{algorithmic}[1] 
\Require Graph $G$, MNI threshold $\tau$, error threshold $\epsilon$
\Ensure A minimal cover set $\Sigma$ of GEDs in $G$ s.t. $\forall \mbox{ }
\sigma\in \Sigma: mni(\sigma.Q)\geq \tau \wedge e_3(\sigma)\leq \epsilon$.
\State{$\mathcal{Q}\gets {\sc GraMi}(G, \tau)$}\Comment{mine freq. graph patterns}\label{l:mfp}
\State{$\mathcal{Q}'\gets reduce(\mathcal{Q})$}\Comment{eliminate redundancies}\label{l:rrp}
\For{$Q\in \mathcal{Q}'$}\label{l:md1}
    \State{$H(Q,G)\gets match(Q,G)$}\Comment{find matches}
    \State{$\Sigma_Q\gets mineDep(H(Q,G), \epsilon)$}\Comment{find appr. GEDs}
    \State{$\Sigma = \Sigma\cup \Sigma_Q$}
\EndFor\label{l:md2}
\State{\Return $\Sigma$}
\end{algorithmic}
\end{algorithm}

\begin{lemma}[GED Characterisation]\label{lem:diff}
Given the itemset $X\subseteq \mathcal{I}$ and an item $\alpha[u;c]\notin X$
over $Q[\bar{u}]$, $X\to \alpha[u;c]$ holds over $H$ iff: $X$ covers 
$nec(\alpha[u;c])$.
$\square$
\end{lemma}

Thus, we reduce the minimal cover of GEDs discovery for a pattern $Q[\bar{u}]$
to the problem of finding the minimal covers for all $nec(\alpha[u;c])$, for 
every item $\alpha[u;c]\in \mathcal{I}$.

\begin{figure}[!t]
    \centering
    \includegraphics[width=0.48\textwidth]{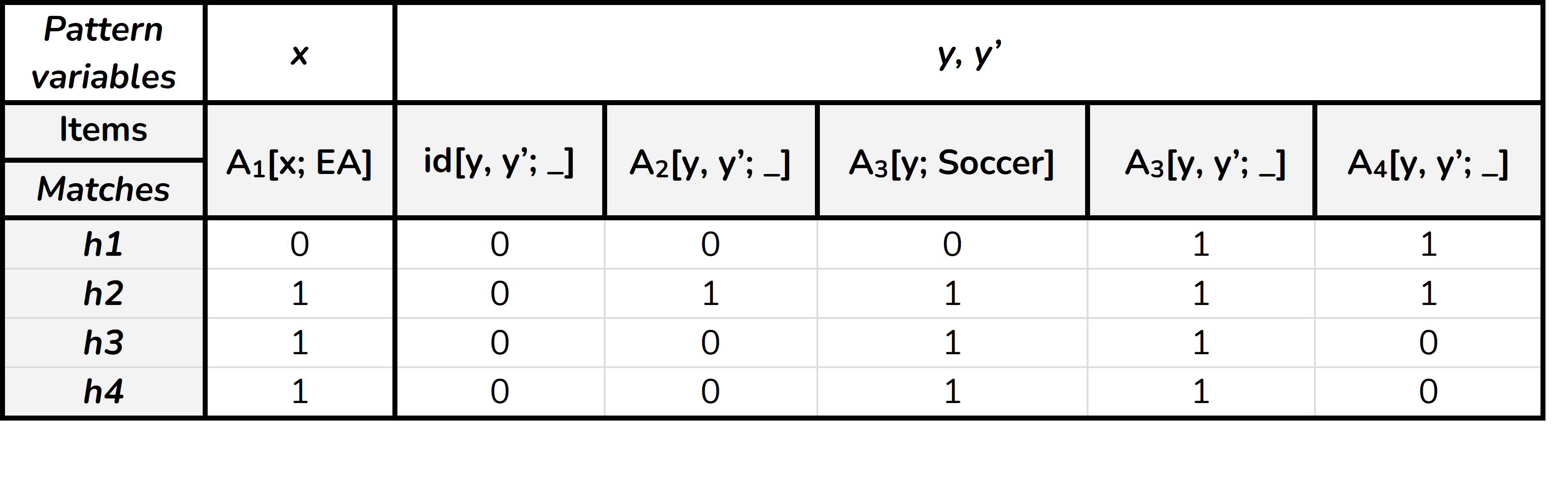}
    \vspace{-10pt}
    \caption{An example disagree table of items over $Q_3$}
    \label{fig:br}
\end{figure}

\smallpara{Construction of Disagree Sets. }
From the foregoing, a fast generation of the disagree and necessary sets is  
crucial. We construct a binary relation $\mathcal{B}(\mathcal{I}, H)$ of matches $H$ over 
the set $\mathcal{I}$ of items over pattern variables in $Q[\bar{u}]$, based on 
the pseudo-relation of $H$.

Given $\mathcal{I}$ and $H$, for each $h\in H$ and $\alpha[u;c]\in \mathcal{I}$
\begin{align}\label{eq:diff}
    \mathcal{B}(h(\alpha[u;c])) := 
    \begin{cases}
        1, & \text{if } h\models \alpha[u;c]\\
        0, & \text{otherwise.}
    \end{cases}
\end{align}
We construct the disagree set $D(h)$ for all $h\in H$, and consequently, the 
necessary set $nec(\alpha[u;c])$ for all items $\alpha[u;c] \in \mathcal{I}$ 
with a single scan of the relation $\mathcal{B}(\mathcal{I}, H)$.

\begin{example}[Disagree \& Necessary Sets]\label{ex:diff}
Using the running example, the binary relation of Fig.~\ref{fig:psuedo-table}
is presented in Fig.~\ref{fig:br}, for a select set of items, using
Equation~\ref{eq:diff}.
The ensuing disagree sets of the matches are as follows\footnote[5]{for brevity,
we represent items by their lowercase letters, e.g. $a_1$ is $A_1[x;EA]$,
$a_3^1$ is $A_3[y;Soccer]$}:
\[
    D(h_1) = \{a_1,id,a_2,a_3^1\}; \qquad D(h_2) = \{id\};
\]
\[
    D(h_3) = \{id,a_2,a_4\}; \qquad D(h_4) = \{id,a_2,a_4\}.
\]
And, the necessary sets of the items are:
\[
    D(id) = \{\{a_1,a_2,a_3^1\}, \{a_2,a_4\}\}; \qquad D(a_1) = \{\{id,a_2,a_3^1\}\};
\]
\[
    D(a_2) = \{\{a_1,id,a_3^1\}, \{id,a_4\}\}; \qquad D(a_3^1) = \{\{a_1,id,a_2\}\}
\] 
\[D(a_4) =\{\{id,a_2\}\}.\square\]
\end{example}

\smallpara{Search Strategy. }
The space of candidate itemsets for a given item $\alpha[u;c]$ is 
$P(\mathcal{I}\setminus\alpha[u;c])$,
which is clearly exponential to the size $|\mathcal{I}|$ (i.e., \(2^{|\mathcal{I}|-1}\)).
Thus, the necessary set, $nec(\alpha[u;c])$, reduces the candidate left
hand sides (LHSs) of $\alpha[u;c]$.
Specifically, to find all valid LHSs of a given item $\alpha[u;c]$, it suffices to
find the set of all covers of $nec'(\alpha[u;c])$, where:
\[nec'(\alpha[u;c]) = \{D\in nec(\alpha[u;c]) \mid \nexists \mbox{ } D'\subset D \in nec(\alpha[u;c])\}.\]

We construct and traverse the lattice of candidate itemsets for a given item $\alpha[u;c]$ 
with its minimal necessary set $nec'(\alpha[u;c])$ using a depth-first, left-to-right strategy. 
Each node in the lattice consists of two elements: the necessary set (enclosed in ``$\{\}$"), 
and its constituent items (in ``$[]$").
We order constituent items at a node in decreasing order of appearance in the necessary itemsets,
and break ties lexicographically.
The root node of the search tree is set to $nec'(\alpha[u;c])$. Each node has up to the number
of unique items in its constituent set, and an edge to a child node is labelled by its corresponding
item. The necessary set of a child node is the set of itemsets in $nec'(\alpha[u;c])$ {\em not}
covered thus far.

There are three possible cases of leaf nodes: a) both node elements are empty;
b) only the remainder necessary set is empty; and c) only the constituent items set is empty.
The resulting branches of these cases correspond to valid LHS, non-minimal LHS,
and invalid LHS respectively.

\begin{figure}
    \centering
    \includegraphics[width=0.48\textwidth]{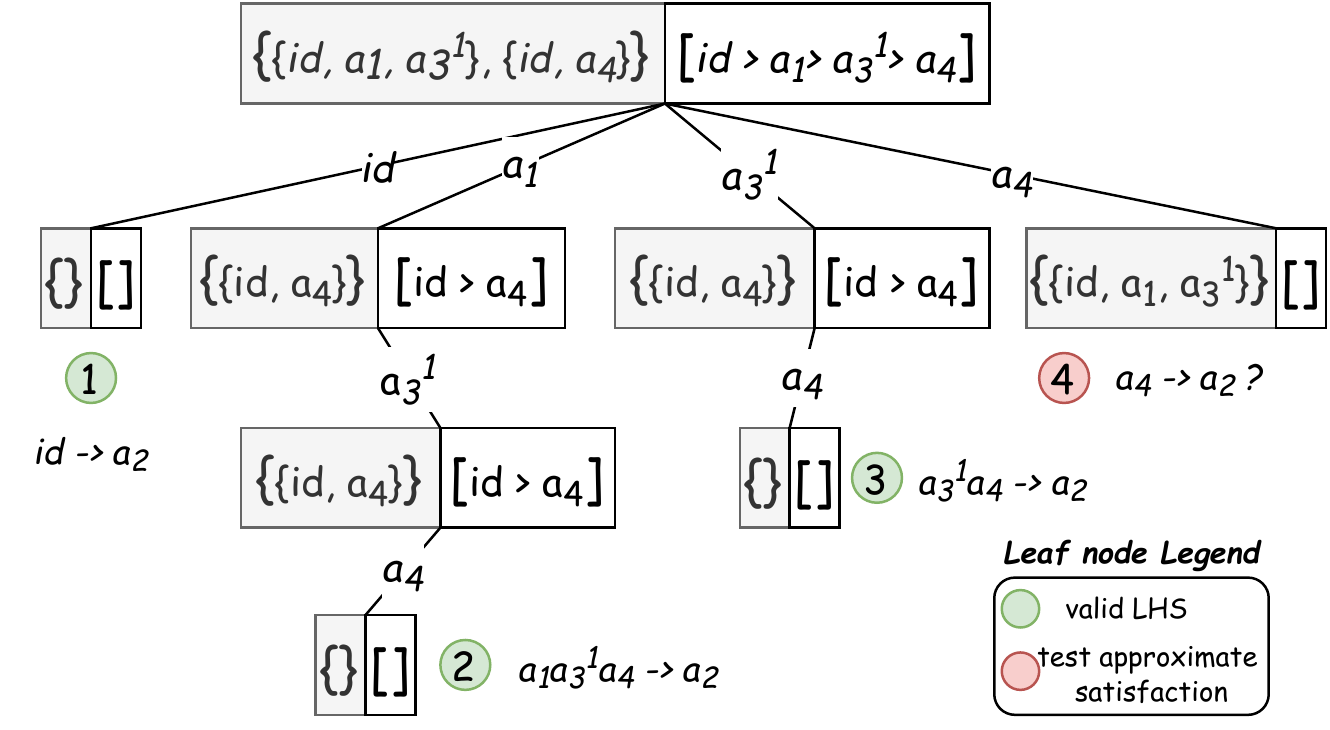}
    \caption{{\sc FastAGEDs} DFS Lattice for $A_2[y,y;\_]$}
    \label{fig:dsf}
\end{figure}

\begin{example}[Find Minimum Cover]\label{ex:mincover}
    The DFS tree for candidate LHSs for $A_2[y,y;\_]$ (i.e., $a_2$) is 
    presented in Fig.~\ref{fig:dsf}. The leaf nodes \(1,2,3\) produce
    valid GEDs with full satisfaction, whereas leaf node \(4\) results
    in an invalid GED. Thus, we test such nodes for approximate satisfaction 
    based on the specified error threshold.
    For instance, let $\epsilon=0.25$; we compute $e_3(Q_3[x,y,y'], a_4\to a_2) =\frac{1}{2}$.
    Thus, leaf node \(4\) does {\em not} produce an approximate GED -- unless 
    $\epsilon\geq 0.5$.
\end{example}

\begin{figure*}
    \centering
    \includegraphics[width=0.96\textwidth]{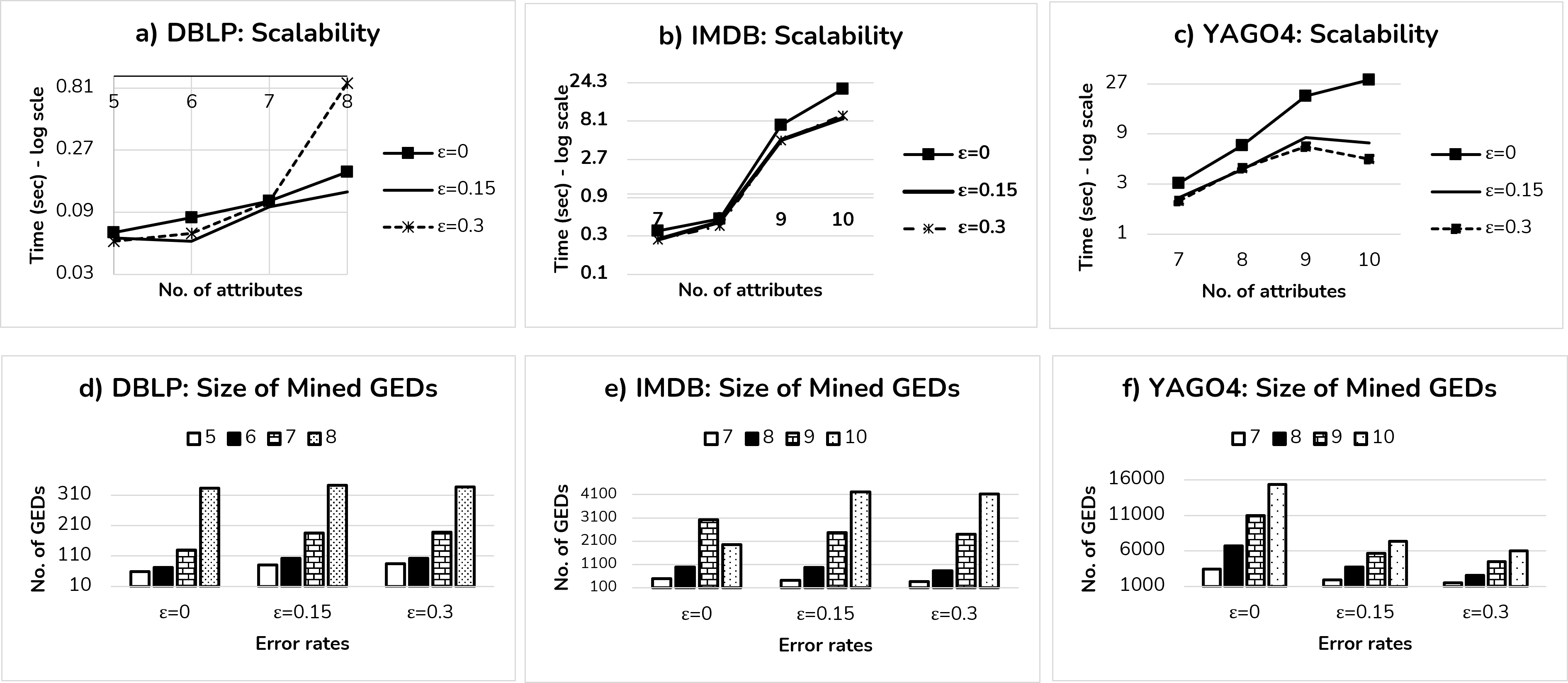}
    \caption{{\sc FastAGEDs}: Experimental Results}
    \label{fig:exp}
\end{figure*}

\section{Experiments}
We next present the experimental study of our algorithm {\sc FastAGEDs} for discovering minimal GEDs. We conducted experiments under different error threshold and numbers of attributes to investigate the effects on the scalability and the number of minimal GEDs mined.
\subsection{Experimental Settings}
\smallpara{Data Sets. }
The experiments were conducted on three real-world property graphs: (a) \emph{IMDB}\cite{imdb} (b) \emph{YAGO4}\cite{yago4} (c) \emph{DBLP}\cite{dblp}. 
Table~\ref{tab1} summarises the details of the data sets.

\begin{table}[!t]
\caption{Summary of the data sets}
\begin{center}
\begin{tabular}{|c|c|c|c|c|}
\hline
Data Set & \#Nodes & \#Edges & \#Node Types & \#Edge Types \\
  \hline
  DBLP & 300K & 800K & 3 & 3  \\
  \hline
  IMDB & 300K & 1.3M & 5 & 8  \\
  \hline
  YAGO4 & 4.37M & 13.3M & 7764 & 83  \\
\hline
\end{tabular}
\label{tab1}
\end{center}
\end{table}

\smallpara{Experiments. }
All the proposed algorithms in this work are implemented in Java; 
and the experiments were run on a 2.20GHz Intel Xeon processor 
computer with 128GB of memory running Linux OS. 

\subsection{Feasibility \& Scalability of Proposal.}
We set the MNI threshold $\tau=2$ in all experiments.

\smallpara{Exp-a. }
We first evaluate the time performance of our algorithm for mining 
(approximate) GEDs in the three datasets, at different error rates 
(i.e., \(\epsilon=0.0,0.15,0.30\)) and for varying number of attributes. 
The results are presented in Fig.~\ref{fig:exp}~a) -- c)
for the {\em DBLP, IMDB} and {\em YAGO4} datasets respectively. 
The results indicate that the time cost increases considerably as the number of attributes increases, regardless of the maximum error we set. Additionally, GED discovery in the DBLP data is the most efficient as it produces the least number of matches for its frequent patterns compared to the other data sets

\smallpara{Exp-b.}
We next report the correlations between the error threshold and the number 
of mined GEDs for differing number of attributes (cf.~Fig.~\ref{fig:exp}d) -- f)). 
In this group of experiments, we set a series of thresholds to conduct this experiments, 
with a range of threshold values from 0 to 0.3.
In general, for a given error rate, the number of mined rules increase with 
increasing attribute size. However, the effect of an increased error rate
for a fixed attribute size in any dataset differs. In exception of the {\em YAGO4}
data where there is a visible decreasing effect, the case is nuanced in the other 
datasets. The phenomenon in the {\em YAGO4} datasets can be explained by its
erroneousness -- it is more dirty than the other datasets. Indeed, an increased 
error-rate returns more rules in principle, however, subsequent implication analysis
results in pruning of more redundant rules. Thus, the resultant effect is a more 
succinct rule set.

\section{Conclusion}
This paper introduces and studies the approximate discovery problem of GEDs.
We propose a measure of error based on the semantics of GEDs, and 
characterise the satisfaction via the concepts of necessary and
cover sets. The developed algorithm, {\sc FastAGEDs}, uses a 
depth-first search strategy to traverse the candidate space of
GEDs to find both full and approximate GEDs. Finally, we show that
{\sc FastAGEDs} is effective and scalable via an empirical 
evaluation over three real-world graphs.


\end{document}